\documentstyle[twoside,amsbsy]{article}
\def\vec#1{{\boldsymbol #1}}
\begin{document}
{\Large\bf 
Bose-Einstein correlations in thermal field theory
}\footnote{\makeatletter
Work supported by GSI, 
hypertext link http://www.gsi.de/gsi.html\\
P.Henning@gsi.de, blasone@tpri6c.gsi.de,razumov@tpri6l.gsi.de}
\\[3mm] \hspace*{6.327mm}\begin{minipage}[t]{12.0213cm}{\large 
P.A.Henning${}^{a,b}$
Ch.H\"olbling${}^{a}$, M.Blasone${}^{a,c}$
and L.V.Razumov${}^{a}$
}\\[2.812mm]
${}^{a}$ Gesellschaft f\"ur Schwerionenforschung GSI\\
         P.O.Box 110552, D-64220 Darmstadt, Germany\\[1mm]
${}^{b}$Institut f\"ur Kernphysik der TH Darmstadt,\\
         Schlo\ss gartenstra\ss e 9, D-64289 Darmstadt, Germany\\[1mm]
${}^{c}$Dipartimento di Fisica dell'
         Universit\`a and INFN,\\ Gruppo Collegato, I-84100 Salerno,
         Italy
\\[4.218mm]{\it 
GSI-Preprint 95-41, \today
}\\[5.624mm] 
{\bf Abstract.} 
Two-particle correlation functions are calculated for bosons emitted from a 
localized thermal source (the ``glow'' of a ``hot spot''). 
In contrast to existing work, non-equilibrium effects
up to first order in gradients of the particle distribution
function are taken into account. The spectral width of the
bosons is shown to be an important quantity: If it is too small,
they do not equilibrate locally and therefore strongly increase 
the measured correlation radius.\\[2mm]
In memoriam of Eugene Wigner and Hiroomi Umezawa.
\end{minipage}

\section*{\large\bf 1. Introduction}
In the field of relativistic heavy-ion collisions
the analysis of Bose-Einstein correlations \cite{HBT56} has attracted much 
attention recently. The general hope is to extract
information about the size of a source radiating mesons by studying
their two-particle correlation function \cite{NA44,NA49}. 
These correlations are typical quantum effects, 
hence quantum field theory is a proper framework 
to describe the problem theoretically. Such efforts have been 
undertaken with great success for many years 
\cite{KP72,GKW79,P84,DS92}, with a big emphasis on 
the quantum properties of mesons that were
emitted from a distribution (in space and time) of classical currents.
The merger of these theoretical considerations with
space-time distributions of quasi-particles generated
by simulation codes for relativistic heavy ion collisions
has led to predictions of correlation radii that are
in rough agreement with experimental data.

The problem of a free quantum field radiating from
a classical current is exactly solvable \cite[pp.438]{IZ80}, also
at finite temperature \cite{h89coh}.
However, in the original field where Bose-Einstein correlations have been used
to measure source radii, i.e., in the Hanbury-Brown--Twiss analysis
of star''light'', as well as in the analysis of
relativistic heavy-ion collisions, the radiation source is a 
localized {\em thermal\/} distribution. Such a state is
a non-equilibrium state, hence somewhat difficult to handle theoretically.

As was pointed out above, 
the usual way to circumvent this problem is the approximation of 
a non-equilibrium state as the superposition of an infinite
number of classical currents, with a certain current distribution
in space and time (mostly assumed to be gaussian) \cite{GKW79}.
From the viewpoint of quantum field theory this approximation
is somewhat unsatisfactory, if not questionable: True {\em non-equilibrium\/}
effects on the propagating particles are neglected, only
{\em local\/} equilibrium effects are maintained.

Since also in relativistic heavy-ion collisions the 
radiating system is off equilibrium, one question arises immediately:
Are there non-equilibrium effects on
the quantum Bose-Einstein correlation ? A secondary
question is, whether the correlation effects generated by
a distribution of classical currents and those generated by a thermal
source are equivalent or not.

To address these questions we consider a stationary temperature
distribution with limited spatial extension, and calculate the two-particle
correlation function of bosons emitted thermally from such a distribution.
In other words, we are investigating correlations in the 
``glow'' from a hot spot -- in close analogy to the physical
situation present in astrophysics, but on the quantum level. 

To this end we
formulate the two-particle correlation problem in a field
theoretical method suited to handle non-equilibrium states.
However, since the solution of the full problem is beyond our
capabilities, we restrict ourselves to the same order of accuracy
that is reached in standard transport theory: Our correlation function
incorporates non-equilibrium effects beyond the local
equilibrium function, but only up to first order in the
gradients of the ``temperature'' distribution of this system.
Furthermore, the discussion is limited to a static non-equilibrium
system.

In another sense we treat the correlation problem 
more consistently than in standard transport theory: We  
take into account a nonzero spectral width for the particles we consider.
This is necessary, because at nonzero temperature {\em every\/}
excitation acquires a finite lifetime due to collisions
with the medium \cite{L88}. We describe this finite lifetime by attributing
a certain spectral width $\gamma>0$ {\em inside the hot spot\/} 
also to asymptotically stable particles.
For strongly interacting particles, like e.g. pions, we may assume
that such a spectral width is due to the coupling 
to $\Delta_{33}$-resonances. 

Quantum field theory for non-equilibrium states comes in two flavors:
The Schwinger-Keldysh method \cite{SKF} and thermo field dynamics (TFD,
see \cite{h94rep,Ubook} for references to the original work). 
For the purpose of the present paper, we prefer the latter method: 
The problem 
of an inhomogeneous temperature distribution has been solved 
explicitly in TFD up to first order in the temperature gradients 
\cite{h94rep}. This solution includes a nontrivial spectral
function of the quantum field under consideration. It employs 
a perturbative expansion in terms of {\em generalized free fields\/}
with continuous mass spectrum \cite{L88}.

The paper is organized as follows. The next section contains a
brief introduction to the formalism
of thermo field dynamics for spatially inhomogeneous systems.
In section 3 we derive expressions for the two-boson
correlation function in non-equilibrium states as well as
its local equilibrium approximation. Section 4 contains a study
of the effect that is exerted on the correlation function by a 
nonzero spectral width, section 5 contains an example with
semi-realistic pion spectral function in hot nuclear matter.
Finally we draw some conclusions and discuss the experimental
relevance.
\section*{\large\bf 2. Outline of the method} 
In ''ordinary'' quantum mechanics,
a statistical state of a quantum system is described by a statistical
operator (or density matrix) $W$, and the measurement of
an observable will yield the average
\begin{equation}\label{av1}
\left\langle \vphantom{\int} {\cal E}(t,\vec{x}) \right\rangle =
   \frac{\mbox{Tr}\left[ {\cal E}(t,\vec{x})\;W \right]}{
         \mbox{Tr}\left[ W \right]}
\;,\end{equation}
where the trace is taken over the Hilbert space of the quantum system
and ${\cal E}$ is the hermitean operator associated with the
observable. In thermo field dynamics (TFD), the calculation
of this trace is simplified to the calculation of a matrix element
\begin{equation}\label{av3}
\left\langle \vphantom{\int} {\cal E}(t,\vec{x}) \right\rangle =
\frac{ (\mkern-4mu( 1 |\mkern-2.5mu| \;{\cal E}(t,\vec{x}) \;
 |\mkern-2.5mu| W)\mkern-4mu)}{
       (\mkern-4mu( 1 |\mkern-2.5mu|  W)\mkern-4mu)}
\;,\end{equation}
with ''left'' and ''right'' statistical state defined in terms of
the two different commuting representations (see refs.
\cite{Ubook,h94rep} for details).

In this state, we consider a complex, scalar boson field
describing spinless charged excitations in a statistical
system not too far from equilibrium. In the spirit of the
first remark, one could think of this field as describing
positive and negative pions in nuclear matter.

According to the reasoning above,
this ``thermal'' boson field is described by two field
operators $\phi_x$, $\widetilde{\phi}_x$  and their adjoints
$\phi_x^\star$, $\widetilde{\phi}_x^\star$, with canonical
commutation relations
\begin{eqnarray}
\left[\phi(t,\vec{x}) , \partial_t
\phi^\star(t,\vec{x}^\prime)\right] & = &
        \mathrm{i}\delta^3(\vec{x}-\vec{x}^\prime) \nonumber \\
\left[\widetilde{\phi}(t,\vec{x}) ,
\partial_t\widetilde{\phi}^\star(t,\vec{x}^\prime)\right] & =-&
         \mathrm{i}\delta^3(\vec{x}-\vec{x}^\prime)
\;\end{eqnarray}
but commuting with each other.
These two fields may be combined in a statistical doublet,
see ref. \cite{h94rep,h93trans} for details.

The free as well as the interacting scalar field
can be expanded into momentum eigenmodes
\begin{eqnarray}\label{bf1}
\phi_x & = &
   \int\!\! \frac{d^3\vec{k}}{\sqrt{(2\pi)^3}}
   \left( a^\dagger_{k-}(t)\,{\mathrm e}^{-\mathrm{i}\vec{k}\vec{x}} +
          a_{k+}(t)\,        {\mathrm e}^{ \mathrm{i}\vec{k}\vec{x}}\right) 
                \nonumber \\
\widetilde{\phi}_x & = &
   \int\!\! \frac{d^3\vec{k}}{\sqrt{(2\pi)^3}}
   \left( \widetilde{a}^\dagger_{k-}(t)\,{\mathrm e}^{\mathrm{i}\vec{k}\vec{x}} +
          \widetilde{a}_{k+}(t)        \,{\mathrm e}^{-\mathrm{i}\vec{k}\vec{x}} 
                \right)
\;.\end{eqnarray}
$\vec{k}$ is the three-momentum of the modes, therefore in this notation
$a^\dagger_{k-}(t)$ creates a negatively charged excitation with momentum
$\vec{k}$, while $a_{k+}(t)$ annihilates a positive charge.
Henceforth the two different charges are distinguished
by an additional index $l=\pm$ whenever possible.

For the free case the commutation relations of the $a$-operators 
at different times are simple, while they are unknown for the 
interacting fields. However, we want to go only one step
beyond the free field case, i.e., we
approximate the fully interacting quantum fields
by {\em generalized free fields\/} \cite{L88}. In this 
formulation, the operators $a$,$\widetilde{a}$ do not excite  
stable on-shell pions. Rather, they are
obtained as an integral over more general operators
$\xi$, $\widetilde{\xi}$ with a continuous energy parameter $E$
\cite{NUY92,h93trans}: 
\begin{eqnarray}\label{bbg4}\nonumber
\left({\array{r} a_{kl}(t)\\
          \widetilde{a}^\dagger_{kl}(t)\endarray}\right)\;=
& \int\limits_0^\infty\!\!dE\,\int\!\!d^3\vec{q}
  \;{\cal A}^{1/2}_l(E,\vec{k})\,
  \left({\cal B}^{-1}_l(E,\vec{q},\vec{k})\right)^\star
  \,\left({\array{r}\xi_{Eql}\\
              \widetilde{\xi}^\#_{Eql}\endarray}\right)
  \,{\mathrm e}^{-\mathrm{i} Et} \nonumber \\
\left({\array{r}a^\dagger_{kl}(t)\\
         -\widetilde{a}_{kl}(t)\endarray}\right)^T\;=
& \int\limits_0^\infty\!\!dE\,\int\!\!d^3\vec{q}
  \;{\cal A}^{1/2}_l(E,\vec{k})\,
  \left({\array{r}\xi^\#_{Eql}\\
           -\widetilde{\xi}_{Eql}\endarray}\right)^T\,
  {\cal B}_l(E,\vec{q},\vec{k})
  \,{\mathrm e}^{\mathrm{i} E t}
\;,\end{eqnarray}
where ${\cal B}$ is a $2\times 2$ matrix, 
the weight functions ${\cal A}_l(E,\vec{k})$ are positive and have
support only for positive energies, their normalization is
\begin{equation}\label{norm}
\int\limits_0^\infty\!\!dE\,E\, {\cal A}_l(E,\vec{k}) =\frac{1}{2}
\;\;\;\;\;\;
\int\limits_0^\infty\!\!dE \, {\cal A}_l(E,\vec{k}) =Z_{kl}
\;.\end{equation}
The principles of this expansion have been discussed in ref. \cite{L88},
its generalization to non-equilibrium states was introduced
in ref. \cite{h94rep}. For equilibrium states the combination
\begin{equation}\label{spec}
{\cal A}(E,\vec{k}) = {\cal A}_+(E,\vec{k})\Theta(E) -
                        {\cal A}_-(-E,-\vec{k})\Theta(-E)
\;\end{equation}
is the spectral function of the
field $\phi_x$
and the limit of free particles with mass  $m$ is recovered when
\begin{equation}\label{fb}
{\cal A}(E,\vec{k}) \longrightarrow
\mbox{sign}(E)\,
  \delta(E^2 -\vec{k}^2 -m^2)
=\mbox{sign}(E)\,
  \delta(E^2 -\omega_k^2)
\;.\end{equation}
For non-equilibrium systems, the existence
of a spectral decomposition cannot be guaranteed \cite{Ubook}.
We may expect however, that close to equilibrium the field 
properties do not change very
much. Thus, with this formalism we study a
quantum system under the influence of small gradients in the
temperature, with {\em local\/} 
spectral function ${\cal A}(E,\vec{k})$. Corrections to such
a picture only occur in second order of temperature gradients
\cite{h93trans,h94rep}. 

A thorough discussion of the $2\times 2$ Bogoliubov matrices was
carried out in ref. \cite{h94rep}. For the purpose of the present paper,
we simply state their explicit form as
\begin{equation}\label{gb}
{\cal B}_l(E,\vec{q},\vec{k}) = \left( { \array{lr}
   \left(\delta^3(\vec{q}-\vec{k}) + N_l(E,\vec{q},\vec{k})\right)
            \;\;\;& -N_l(E,\vec{q},\vec{k}) \\
   -\delta^3(\vec{q}-\vec{k})     & \delta^3(\vec{q}-\vec{k})
   \endarray} \right)
\;,\end{equation}
where $N(E,\vec{q},\vec{k})$ is the Fourier transform of a space-local
Bose-Einstein distribution function
\begin{eqnarray}\nonumber \label{nloc}
N_l(E,\vec{q},\vec{k})& = & \frac{1}{(2\pi)^3}\;
        \int\!\!d^3\vec{z}\,{\mathrm e}^{-\mathrm{i}
  (\vec{q}-\vec{k})\vec{z} }\,n_l(E,\vec{z})\\
n_l(E,\vec{z}) & = &  
  {\displaystyle \frac{1}{{\mathrm e}^{\beta(\vec{z}) (E-\mu_l(\vec{z}))}-1}}
\;.\end{eqnarray}
Here we have assumed a distribution function that only depends on the
energy parameter $E$ and on the space coordinate $z$. The expansion
allows for a generalization of this, to more general distribution functions
depending also on the momentum $(\vec{q}+\vec{k})/2$.

We have argued, that our ansatz for the fields gives rise to a local
spectral function. A moving particle however feels an influence also of
the {\em gradients\/} of this local equilibrium distribution.
Consequently also the propagator for the fields we consider is correct beyond
a local equilibrium situation, to be precise it is correct to first order 
in the gradients of $n_l(E,\vec{z})$.

To complete the brief description of the TFD formalism, we specify the
commutation relation of the various operators in our expressions.
The $\xi$-operators have commutation relations
\begin{equation}\label{difc}
\left[\xi_{Ekl},\xi^\#_{E^\prime k^\prime l^\prime}\right]=
  \delta_{ll^\prime}\,
  \delta(E-E^\prime)\,
\delta^3(\vec{k}-\vec{k}^\prime)
\;.\end{equation}
Similar relations hold for the $\widetilde{\xi}$ operators, 
all other commutators vanish, see \cite{L88}. 
It follows from these definitions, that
\begin{eqnarray}\label{co}
\left[a_{kl}(t),a^\dagger_{k^\prime l^\prime}(t)\right]&=&
  Z_{kl}\,
  \delta_{ll^\prime}\,
  \delta^3(\vec{k}-\vec{k}^\prime)\nonumber \\
\left[\widetilde{a}_{kl}(t),
  \widetilde{a}^\dagger_{k^\prime l^\prime}(t)\right]&=&
  Z_{kl}\,
  \delta_{ll^\prime}\,
  \delta^3(\vec{k}-\vec{k}^\prime)
\; \end{eqnarray}
are the equal-time commutation relations for the $a$, $\widetilde{a}$
operators.

The $\xi$, $\widetilde{\xi}$ operators act on the
''left'' and ''right'' statistical state according to
\begin{equation}\label{tscc}
\xi_{Ekl}|\mkern-2.5mu| W )\mkern-4mu) = 0, \;\;
\widetilde{\xi}_{Ekl}|\mkern-2.5mu| W )\mkern-4mu) =0,\;\;
(\mkern-4mu( 1|\mkern-2.5mu| \xi^\#_{Ekl} = 0 ,\;\;
(\mkern-4mu( 1|\mkern-2.5mu| \widetilde{\xi}^\#_{Ekl} = 
0\;\;\;\forall\,E,\vec{k},l=\pm1
\;. \end{equation}
With these rules, all bilinear expectation values can be
calculated exactly. Higher correlation functions have
a perturbative expansion in the spectral function.
\section*{\large\bf 3. Two-particle correlation function}
Of the higher correlation functions, 
we are interested in the two-particle correlation function,
which is the probability to find 
in the system a pair of pions with momenta $p$ and $q$. 
For the non-equilibrium system we are considering, this
correlation function is
\begin{equation} \label{cfc}
c_{ll^\prime}(\vec{p},\vec{q})=\frac{\left\langle \vphantom{\int}
 a^\dagger_{pl}(t) 
 a^\dagger_{ql^\prime}(t)
            a_{ql^\prime}(t) a_{pl}(t)\right\rangle}{
   \left\langle \vphantom{\int}a^\dagger_{pl}(t) a_{pl}(t)\right\rangle\;
   \left\langle \vphantom{\int}a^\dagger_{ql^\prime}(t) a_{ql^\prime}(t)
   \right\rangle}
=1+\delta_{ll^\prime}
\frac{{\cal F}(\vec{p},\vec{q}){\cal F}(\vec{q},\vec{p})
      }{{\cal F}(\vec{p},\vec{p}) {\cal F}(\vec{q},\vec{q})}
\;.\end{equation}
For simplicity, we abbreviate the mean momentum of this pair by
$\vec{Q}= (\vec{q}+\vec{p})/2$. The function 
${\cal F}(\vec{q},\vec{p})$ is calculated using the 
standard rules of thermo field dynamics given above. One
obtains
\begin{equation} \label{cfcf}
{\cal F}(\vec{p},\vec{q})  = 
      \int\limits_{0}^{\infty}\!dE\int\!d^3\vec{z}\,
      \left({\cal A}_l(E,\vec{p})
            {\cal A}_l(E,\vec{q})\right)^{\frac{1}{2}}\,
      \mbox{e}^{{\mathrm i}(\vec{p}-\vec{q})\vec{z}}\,
      n_{l}(E,\vec{z})
\;, \end{equation}
where one may also insert a $\vec{z}$-dependent spectral function
without violating the accuracy to first order in the gradients.

How these gradients enter the above expressions may be seen when
performing an expansion of ${\cal A}$ around the mean momentum $\vec{Q}$.
\begin{eqnarray} \label{cfc2} \nonumber
{\cal F}(\vec{p},\vec{q}) & = &  {\cal F}^0(\vec{p},\vec{q}) \\  
\nonumber 
&+&
   \int\limits_{0}^{\infty}\!dE\int\!d^3\vec{z}\,
   \mbox{e}^{{\mathrm i}(\vec{p}-\vec{q})\vec{z}}\,\left(
   {\mathrm i} \nabla_{\vec{Q}}{\cal A}_l(E,\vec{Q},\vec{z})\,
   \nabla_{\vec{z}} n_{l}(E,\vec{z})\right)\\
&+&{\cal O}(\nabla^2_z n)
\;.\end{eqnarray}
Here, the lowest order term 
\begin{equation}\label{cfc3} 
{\cal F}^0(\vec{p},\vec{q}) =  
\int\limits_{0}^{\infty}\!dE\int\!d^3\vec{z}\,
   \mbox{e}^{{\mathrm i}(\vec{q}-\vec{p})\vec{z}}\,
   {\cal A}_l(E,\vec{Q},\vec{z})\,
   n_{l}(E,\vec{z})
\;\end{equation} 
is the {\em local equilibrium\/} contribution to the
${\cal F}$ we have obtained above. 

For a possible generalization, i.e., to explicitly
momentum dependent $n_l$, it is worthwhile to note
that the gradient term in (\ref{cfc2}) is just one half of the 
Poisson bracket of ${\cal A}$ and $n$
\cite{h94rep}. Furthermore, we find that
$
{\cal F}(\vec{p},\vec{p})={\cal F}^0(\vec{p},\vec{p})
$,
i.e., the denominator of the correlation function is not affected by the
gradient expansion.

We therefore obtain as the 
{\em local equilibrium two-particle correlation function\/} the expression
\begin{equation}\label{cfcs}
c_{ll^\prime}^{\mbox{\small loc}}(\vec{p},\vec{q})
  =1+\delta_{ll^\prime}
\frac{{\cal F}^0(\vec{p},\vec{q}){\cal F}^0(\vec{q},\vec{p})
      }{{\cal F}^0(\vec{p},\vec{p}) {\cal F}^0(\vec{q},\vec{q})}
\;.\end{equation}
However, the full {\em non-equilibrium correlation function\/}
$c_{ll^\prime}(\vec{p},\vec{q})$ is the one measured experimentally. 

The exact correspondence between the local equilibrium result and 
other calculations of the correlator 
\cite{KP72,GKW79,P84,DS92} may be found when inserting the 
free spectral function from (\ref{fb}):
\begin{equation}\label{cfc4}
c_{ll^\prime}^{\mbox{\small free}}(\vec{p},\vec{q})
  =1+\delta_{ll^\prime}
\frac{\displaystyle \left|\int\!d^3\vec{z}\,
   \mbox{e}^{{\mathrm i}(\vec{q}-\vec{p})\vec{z}}\,
   n_{l}(\omega_{\vec{Q}},\vec{z})\right|^2}{\displaystyle
\left(\int\!d^3\vec{z}\,
   n_{l}(\omega_{\vec{p}},\vec{z})\right)
\left(\int\!d^3\vec{z^\prime}\,
   n_{l}(\omega_{\vec{q}},\vec{z^\prime})\right)}
\;.\end{equation}
Obviously, one may not insert the free spectral function 
into equation (\ref{cfcf}) for ${\cal F}$. This is
{\em not\/} a flaw of the derivation, but suggests -- as expected -- 
that the limit of zero spectral width at finite temperature is ill-defined
\cite{h94spec}. 
\section*{\large\bf 4. Simple spectral function}
In this section we study the difference between the local equilibrium
correlation function (\ref{cfcs})
and the non-equilibrium result (\ref{cfcf}) in more detail.
To this end we calculate the correlation functions 
with a simple parameterization of a boson (pion) spectral function, 
\begin{equation}\label{sfs}
  {\cal A}_l(E,\vec{p})=\frac{2E\gamma}{\pi}
  \frac{1}{(E^{2}-\Omega_{p}^{2})^{2}+4E^{2}\gamma^{2}}
\end{equation}
where $\Omega_{p}=\sqrt{m_{\pi}^{2}+\vec{p}^{2}+\gamma^{2}}$ and
$m_\pi$ = 140 MeV. To 
gain information about the {\em maximal\/} influence exerted
by the occurrence of a nonzero spectral width, we use
an energy and momentum independent $\gamma$ equal for both charges. 

The temperature distribution is taken as a radially symmetric
gaussian,
\begin{equation} \label{tempd}
 T(\vec{z})=T(r) = T_0 \exp\left(-\frac{r^{2}}{2R_0^{2}}\right)
\;,\end{equation}
with chemical potential $\mu=0$ and $R_0$ = 5 fm.

The local equilibrium pion distribution
for a given momentum $\vec{k}$ is obtained by folding $n(E,\vec{z})$ 
with the
spectral function. Hence, the mean radius of the {\em particle\/}
distribution function acquires a $\gamma$-dependence. We define
the rms radius {\em orthogonal\/} to the direction of $\vec{k}$ as
\begin{equation}\label{rav}\nonumber
R_{\mbox{\small rms}} = \sqrt{\frac{I_2}{I_0}} \;\;\;\;\;\;\;
I_j = \int\limits_0^\infty \!\!dr\,r^j\,\int\limits_0^\infty
  \!\!dE\,{\cal A}(E,\vec{k})\,\left(\mbox{e}^{E/T(r)}-1\right)^{-1}
\;.\end{equation}
Note, that $R_{\mbox{\small rms}}$ is {\em not\/} the 3-dimensional
rms radius of the distribution function (which would be $I_4/I_2$).
Rather, $R_{\mbox{\small rms}}$ is half the product of angular
diameter and distance between detector and source.
A constant temperature over a sphere of radius $R_0$
would yield an $R_{\mbox{\small rms}} = R_0/\sqrt{3}$, while
its 3-D rms radius is $R_0 \sqrt{3/5}$.

\begin{figure}[t]
\setlength{\unitlength}{1mm}
\vspace*{100mm}
\includegraphics{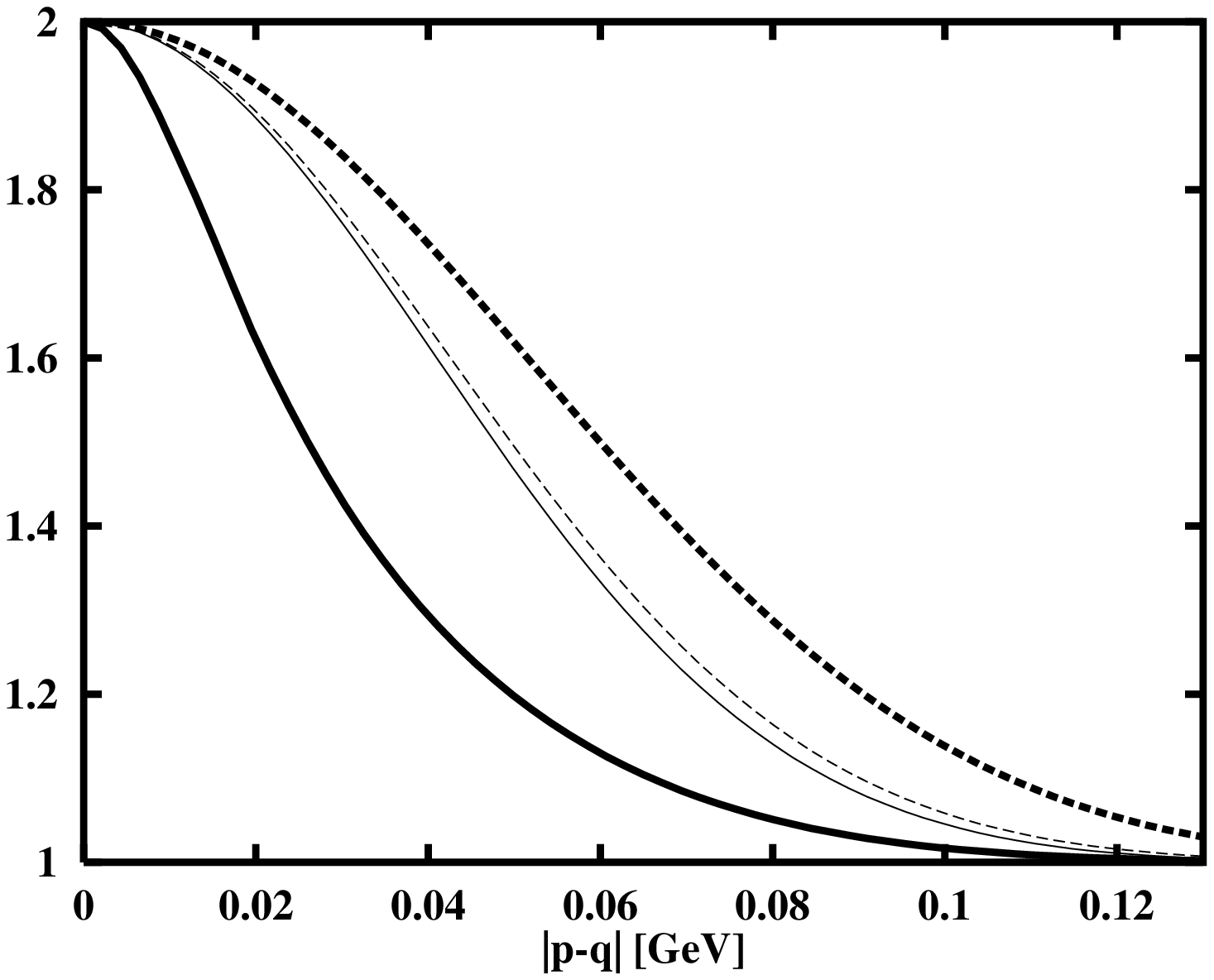}
\caption{Correlation function for a pion ``hot spot'' with 
temperature $T_0$=100 MeV.}
{\small
Simple spectral function from eq. (\ref{sfs}), 
$(\vec{p}+\vec{q})/2$= 100 MeV.\\
Thin lines: $\gamma$=50 MeV, 
Thick lines: $\gamma$=5 MeV.\\
Solid lines: non-equilibrium correlation function (\ref{cfc}).\\
Dashed lines: local equilibrium correlation function (\ref{cfcs}).
}\\\hrule
\label{fig1}
\end{figure}

In fig. \ref{fig1} we have plotted the correlation functions
for two different constant values of the parameter $\gamma$.
Clearly, for $\gamma$=50 MeV the correlation function 
$c_{ll^\prime}(\vec{p},\vec{q})$ agrees quite well with the
local equilibrium result, and very closely resembles a gaussian.

However, for smaller $\gamma$=5 MeV the non-equilibrium correlation function
is much narrower in momentum space than the local equilibrium result, 
it also deviates from a gaussian form. Nevertheless we may
approximate it by such a simple functional form in order to extract 
quantitative information, i.e.,
\begin{equation}\label{gau}
c_{ll^\prime}(\vec{p},\vec{q}) \approx 
1 + \exp\left( -R^2 (\vec{p}-\vec{q})^2 \right)
\;\end{equation}
and similarly for $c_{ll^\prime}^{\mbox{\small loc}}(\vec{p},\vec{q})$ 
with parameter $R_{\mbox{\small loc}}$.
 
\begin{figure}[t]
\setlength{\unitlength}{1mm}
\vspace*{100mm}
\includegraphics{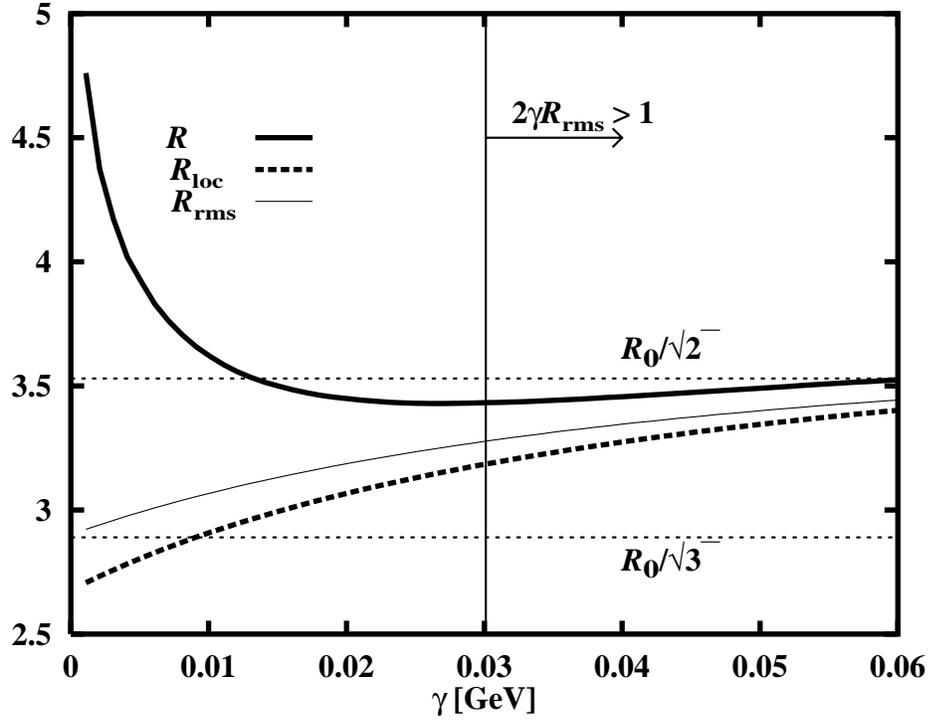}
\caption{Correlation radius of a pion ``hot spot'' with 
temperature $T_0$=100 MeV.}
{\small
Simple spectral function from eq. (\ref{sfs}), 
$(\vec{p}+\vec{q})/2$= 100 MeV.\\
Thick solid line: gaussian fit radius $R$ of the correlation
function (\ref{cfc}). \\
Thick dashed line: gaussian fit radius $R_{\mbox{\small loc}}$ 
of the correlation function (\ref{cfcs}). \\
Thin solid line: $\gamma$-dependent rms radius $R_{\mbox{\small rms}}$.\\
}\\\hrule
\label{fig2}
\end{figure}

In figure \ref{fig2} we show the two fit parameters $R$ and 
$R_{\mbox{\small loc}}$ as function of $\gamma$, together
with the $\gamma$-dependent rms radius of the particle distribution.
We find that the {\em measured\/} correlation radius $R$
is always larger than $R_{\mbox{\small loc}}$, with a minimum
reached at $\gamma\approx 26.7$ MeV.

The plot may be divided in two regions, with a boundary at
$
2\gamma R_{\mbox{\small rms}}[\gamma] = 1 \Leftrightarrow 
\gamma\approx 30.2$ MeV. For these two regions we find 
\begin{equation}
{\array{lllll}
2\gamma R_{\mbox{\small rms}} \;\ll\; 1\;&\;\Rightarrow\;&\;
 R \approx R_{\mbox{\small loc}} + 
  1/\gamma \; &\;\gg\; R_{\mbox{\small rms}}\;&\; >
  \;R_{\mbox{\small loc}}\\
2\gamma   R_{\mbox{\small rms}}\;\gg\; 1\;&\;\Rightarrow  &\;
 R \approx R_{\mbox{\small loc}} 
              &\stackrel{>}{\sim}\; R_{\mbox{\small rms}}& >\;
  \;R_{\mbox{\small loc}}
\;.\endarray}
\end{equation}
For larger $\gamma$, the small differences between $R$, 
$R_{\mbox{\small loc}}$ and $R_{\mbox{\small rms}}$ may be attributed to our 
use of a gaussian temperature distribution:
$n(T(r))$ is not strictly gaussian, only in the (unphysical) limit
$\gamma\rightarrow\infty$ one reaches 
$R=R_{\mbox{\small loc}}=R_{\mbox{\small rms}}=R_0/\sqrt{2}$
 
The interpretation of this result is straightforward:
A finite lifetime 
or nonzero spectral width $\gamma>0$ of the bosons {\em inside the
source\/} is essential,
if one wants to infer the {\em thermal source radius\/} $R_0$ 
from correlation measurements. To be more precise, 
only for $2\gamma R_{\mbox{\small rms}} \ge 1$ 
the correlation function measures 
the mean diameter of the particle distribution function $n(E,\vec{z})$. 

This result is in agreement with our view of the {\em relaxation
process\/} of a non-equilibrium distribution function:
The relaxation rate is, to
lowest order, given by the spectral width of the particle \cite{h94rep}.
Consequently, {\em zero\/} $\gamma$ corresponds to a system 
without dissipation. In such a system, only quantum coherence effects
exist, and thus the correlation function approaches
the ``quantum limit'' 
\begin{equation}\label{qaml}
\lim_{\gamma\rightarrow 0}
c_{ll^\prime}(\vec{p},\vec{q}) = 1 + 
\delta_{ll^\prime} \delta_{\vec{p}\,\vec{q}}
\;.\end{equation}
For this case, 
the correlation radius obtained by a gaussian fit becomes infinite. 
We may also view, for a given energy, $1/\gamma$ as a 
measure for the spatial size of the pion ``wave packet'', which must
be smaller than the object to be resolved. In other words, the
mean free path of the bosons must not exceed the object size to produce
correlations.
\section*{\large\bf 5. Semi-realistic spectral function}\label{real}
To get a more realistic result for the non-equilibrium two-pion correlation
function measured in the thermal radiation from a hot spot
in nuclear matter, we use the the spectral function derived
in \cite{h94rep,h93pion}. It includes the coupling of pions to 
$\Delta_{33}$-resonances in nuclear matter, which are taken to 
have a constant spectral width $\Gamma$ by themselves.

It was argued in ref. \cite{h94rep}, that using such an approximate
spectral width for the $\Delta_{33}$ resonances constitutes
the only way to achieve an analytical solution of the 
$\Delta$-hole polarization problem in nuclear matter. An even
more realistic energy-momentum dependent spectral width for
the $\Delta_{33}$ resonance can be treated only in a fully self-consistent
numerical treatment involving dispersion integral techniques.

To first order in such a constant $\Gamma$, the pionic spectral
function is
\begin{equation}\label{asys}
{\cal A}(E,\vec{k},\vec{z}) =
\frac{\vec{k}^2C(\vec{z})}{\pi}\,\frac{\Gamma\,E\,\omega_\Delta}{
   (E^2-\omega^{\prime\,2}_+)^2\,
   (E^2-\omega^{\prime\,2}_-)^2 +  \Gamma^2\,E^2\,(E^2-E^2_\pi)^2}
\;.\end{equation}
Note, that this function is coordinate dependent. 
The energies in the denominator are 
\begin{equation}\label{omf2}
\omega^{\prime\, 2}_\pm = \frac{1}{2}\left(E_{N\Delta}^2
 +(\Gamma/2)^2+E_\pi^2 \pm
  \sqrt{ \left(E_{N\Delta}^2+(\Gamma/2)^2
-E_\pi^2\right)^2 + 4\vec{k}^2 C(\vec{z}) \omega_\Delta}
  \right)
\;,\end{equation}
with functions
\begin{eqnarray}\nonumber\label{odf2}
\omega_\Delta  & = & E_\Delta(\vec{k}) -M_N
    = \sqrt{\vec{k}^2+M_\Delta^2}-M_N\\
\nonumber
C(\vec{z}) & = &
  \frac{8}{9}\left(\frac{f^\pi_{N\Delta}}{m_\pi}\right)^2\,
  \left( \rho^0_N(\vec{z})-\frac{1}{4}\rho^0_\Delta(\vec{z})\right)\\
E_{N\Delta}(\vec{k}) & = & \vphantom{\int}\sqrt{
     \omega_\Delta(\vec{k})
     \left( \omega_\Delta(\vec{k}) + g^\prime\,C(\vec{z})\right)}
\;.\end{eqnarray}
\begin{table}[t]
\begin{center}
\begin{tabular}{ccccccc} \hline
~$f^\pi_{N\Delta}$~ & ~$g^\prime$~ & ~$m_\pi$~ & ~$M_N$~     &
         ~$M_\Delta$~ & ~$\Gamma$~ & $\rho_0$\\ \hline
         2          & ~0.5~        & ~0.14 GeV~& ~0.938 GeV~ &
          ~1.232 GeV~ & ~0.12 GeV~& 0.155 fm ${}^{-3}$  \\ \hline
\end{tabular}
\end{center}
\caption{Coupling constants and masses used in the calculations of
 this work.}
{\small
The value of $g^\prime$ was chosen to allow for a direct
comparison with simulation codes for heavy-ion collisions,
a more realistic value to describe
pion scattering data would be $g^\prime\approx 1/3$.}\\ \hrule
\end{table}
The baryon number in each small volume and hence the baryon density is a 
constant parameter of the calculations,
for free particles 
with bare ''on-shell'' energies
\begin{equation}
E_N(\vec{p})  =  \sqrt{\vec{p}^2 + M_N^2}\;\;\;\mbox{and}\;\;\;
E_\Delta(\vec{p})  =  \sqrt{\vec{p}^2 + M_\Delta^2}
\;\end{equation}
this baryon density is obtained as
\begin{eqnarray}\nonumber\label{rb0}
\rho_b^0& =& \rho^0_N(\vec{z}) + \rho^0_\Delta(\vec{z})  \\
      & =& 4\,\int\!\!\frac{d^3\vec{p}}{(2\pi)^3}\,n_N(E_N(\vec{p}),\vec{z})
          +16\,\int\!\!\frac{d^3\vec{p}}{(2\pi)^3}\,
      n_\Delta(E_\Delta(\vec{p}),\vec{z})
\;.\end{eqnarray}
The distribution functions are taken as local Fermi-Dirac functions
\begin{equation}
n_{N,\Delta}(E,\vec{z})   =   
 \frac{1}{{\mathrm e}^{(E-\mu_{N,\Delta})/T(\vec{z})}+1}
\;,\end{equation}
with temperature $T(\vec{z})$. 
As temperature distribution we use the same as 
in the previous section, eq. (\ref{tempd}), but with different central 
temperatures $T_0$.
\begin{table}[b]
\hrule
\begin{center}
\begin{tabular}{llllll}
$T$ & $(\vec{p}+\vec{q})/2$ & $R$ & $R_{\mbox{\small loc}}$ &
$R_{\mbox{\small rms}}$ & $R/R_{\mbox{\small rms}}$   \\ \hline
100 MeV & 100 MeV & 3.99 fm & 2.89 fm & 3.14 fm &1.27 \\ 
160 MeV & 100 MeV & 4.86 fm & 3.23 fm & 3.51 fm &1.38 \\  \hline
100 MeV & 350 MeV & 3.23 fm & 3.16 fm & 3.34 fm &0.97 \\ 
160 MeV & 350 MeV & 3.57 fm & 3.49 fm & 3.73 fm &0.96 \\ \hline
\end{tabular}
\end{center}
\caption{Correlation radii obtained by gaussian fit to
the correlation function with semi-realistic spectral function.}
\label{tab2}
\end{table}

Since the temperature depends on the spatial coordinate, 
fixed baryon density implies that the ``baryochemical''
composition of the hot spot changes with coordinate $\vec{z}$.
In the center of the hot spot baryons are, to a large extent,
present in the form of $\Delta$-resonances. Outside the
hot region baryons are ``only'' nucleons -- and as shown before, the
pion properties there do not influence our calculation.

\begin{figure}[t]
\setlength{\unitlength}{1mm}
\vspace*{180mm}
\includegraphics{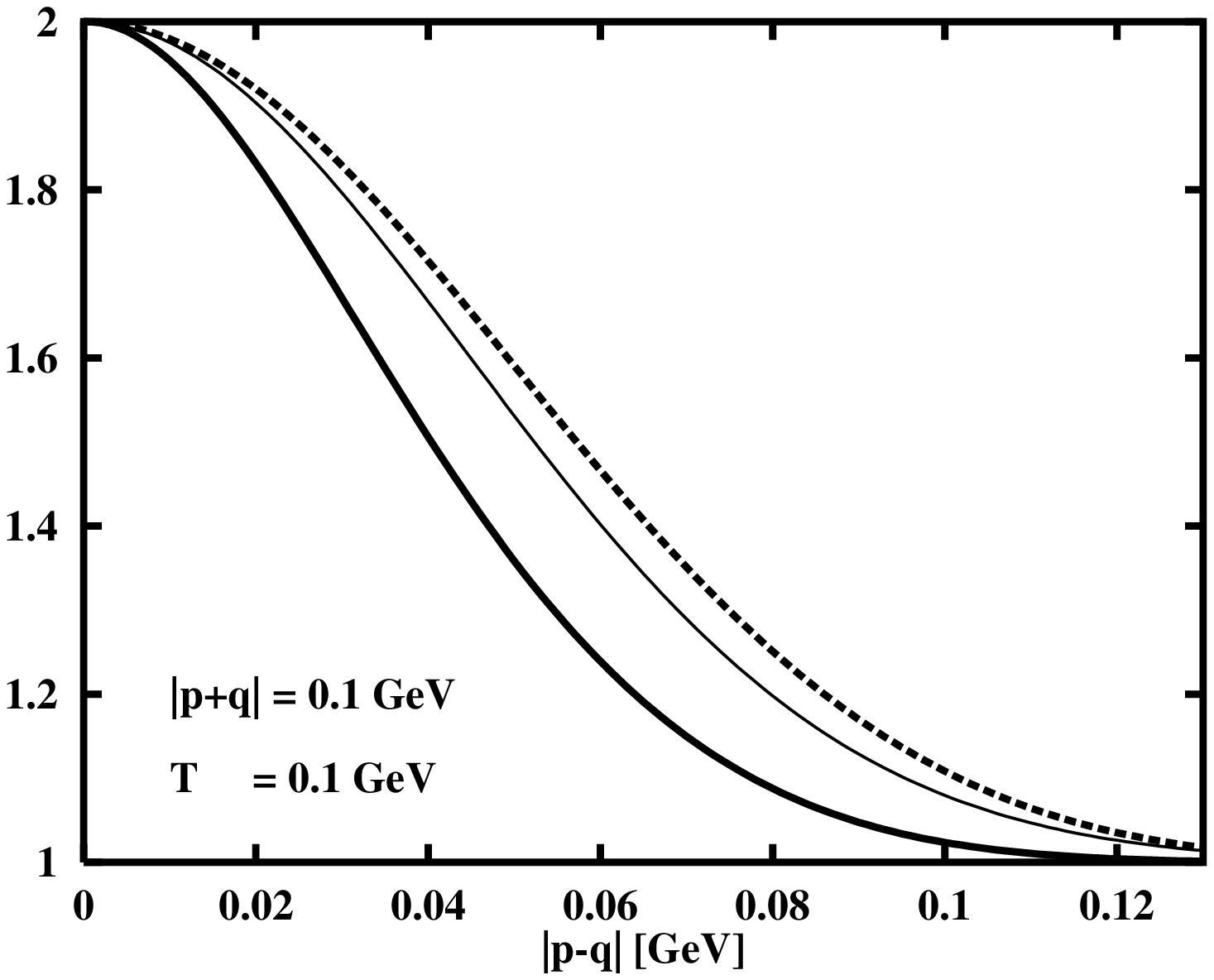}
\includegraphics{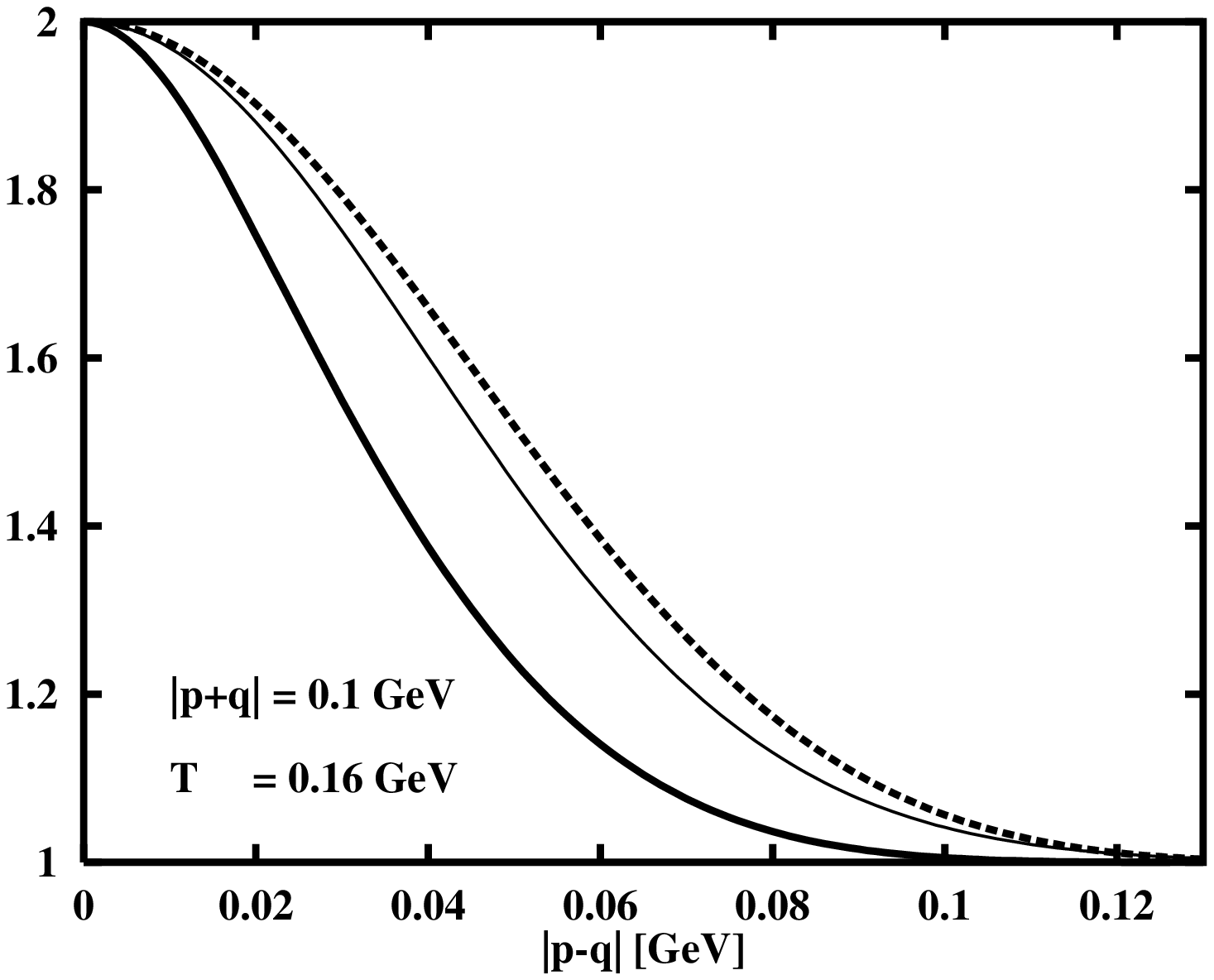}
\caption{Correlation function for pions from a ``hot spot'' with 
different central temperature.}
\label{fig3}
{\small
Semi-realistic spectral function from eq. (\ref{asys}),
$(\vec{p}+\vec{q})/2$= 100 MeV.\\[1mm]
Top panel: Central temperature $T_0$ = 100 MeV,\\
Bottom panel: Central temperature $T_0$ = 160 MeV,\\[1mm] 
Solid thick line: non-equilibrium correlation function (\ref{cfc}).\\
Dashed line: local equilibrium correlation function (\ref{cfcs}).\\
Solid thin line: gaussian with correlation radius equal to
 $R_{\mbox{\tiny rms}}$ (see table).
}\vspace*{1mm}
\end{figure}

In fig. \ref{fig3}, we have plotted the correlation function
for a pair momentum of $\vec{Q}=(\vec{p}+\vec{q})/2 =$ 100 MeV
at the two values $T_0$= 100 MeV and $T_0$ = 160 MeV.
In this momentum region, the spectral width of the pion is small
due to its pseudo-vector coupling with baryons. This we
assume to be a general feature of pions in nuclear matter,
although one may argue about the exact value of $\gamma$. 
We find, that at higher temperature the non-equilibrium correlation
function (which we had assumed to be the one measured experimentally)
is much narrower in momentum space than the local equilibrium
function.

The local equilibrium correlation function
however is close to the the rms-radius of the thermal source.
The correlation radii obtained by a gaussian fit to the 
non-equilibrium as well as the local equilibrium distribution are given 
in table \ref{tab2}. Following these results we conclude, that the
effect we propose is absent at higher momentum of the pions,
where the p-wave coupling to nuclear matter is big enough to
give it a sufficiently large spectral width for local
equilibration. In the low momentum region, the 
{\em measured\/} correlation radius
overestimates the source size by as much as 30 - 40 \%.
\section*{\large\bf 6. Conclusions}
Before we draw a final conclusion from our work, we must emphasize that
it is still too early to use our result for the correlation function in 
a direct comparison to experimental data.
For any realistic situation, we certainly have to take into account
also the partially {\em coherent\/} production of pions:
A substantial fraction of
pions arriving in a detector stems from the free-space decay
of $\Delta_{33}$ resonances, thus forcing 
$c_{ll^\prime}(\vec{p},\vec{p})<2$.

Furthermore it seems worthwhile to note that our
derivation does not contradict existing 
work on the correlation function. In 
an early semiclassical treatment, an infinite lifetime
(equivalent to $\gamma\equiv0$)
of an excitation was found to produce a correlation function $c \equiv 1$
\cite{KP72}, whereas we find an infinitely narrow peak in this
``quantum limit'', see eq. (\ref{qaml}.

Another paper based on the Schwinger-Keldysh method also finds
that the total production rate of particles is proportional to 
an energy integral over the off-diagonal self-energy components,
i.e., to the integral over $\gamma\cdot n$ in our formalism \cite{DS92}.
However, due to their explicit quasi-particle approximation
the authors of ref. \cite{DS92} lose the consistent treatment 
of the non-equilibrium effects
to first order in the temperature gradients.

Let us briefly remark on the physical situation
present in {\em stars\/} emitting photons:
For these we may assume a thermal width of the photon which
is very small, e.g., $\gamma\approx \alpha T\approx 0.5/137$ eV.
However, the source radius $R_{\mbox{\small rms}}$ of a star is, in general,
so big that the condition $2\gamma R_{\mbox{\small rms}} \gg 1$ 
is always satisfied. Hence our results do {\em not\/} affect correlation radii
measured for astronomical objects.

Having in-lined our calculation with existing work, we may now
carefully conclude the following physical effect relevant
for relativistic heavy-ion collisions: 
When pion pairs with a relatively low 
momentum are created in hot nuclear matter, their 
p-wave interaction with the surrounding medium is small.
Hence, in their movement outwards from the hot zone they do
not have a sufficiently short mean free path to be locally
equilibrated. Thus, their correlation length is dominated by
their mean free path ($\approx 1/\gamma$) rather than 
by the thermal source radius.

As we have shown, this leads to a correlation function
which is {\em narrower\/} in momentum space than expected
in a local equilibrium situation: Compare the solid and the
dashed curves in fig. \ref{fig3}. Consequently, measuring
the correlation function of low momentum pions in
a non-equilibrium state 
{\em overestimates\/} the source radius. For a semi-realistic
pion spectral function and a pair momentum of 100 MeV, we find
this effect to be as large as 30 -- 40\%, depending strongly
on the actual source size.

Turning to experimental results of NA44, we learn that
correlation measurements of pions and K-mesons emanating from
relativistic heavy-ion collisions yield comparable fitted
correlation radii of 3--4 fm \cite{NA44}. As we have shown, it
might be premature to conclude from these measurements
that the {\em source size\/} for both mesons is similar: One
would have to compare the mean free path of both particle species before
such a conclusion.

More recently the experiment NA49 has measured pion correlation
functions in central Pb+Pb collisions at 33 TeV. Resulting was
a correlation radius of 6-7 fm, as compared to 4.5 fm
in S+Au collisions \cite{NA49}. It is a particularity of our
semi-realistic pion spectral function that it is narrower for
higher temperature and low momenta. Physically, the first
effect stems from the higher $\Delta_{33}$ abundance in the hotter system: 
$\Delta$-particle/nucleon-hole pairs are less easily polarized.
The second effect is due to the p-wave coupling of pions to nucleons,
as was pointed out above.

Consequently, the effective $\gamma$ of the pions drops with
increasing temperature -- and one measures a higher correlation radius.
If we presume this effect, which was deduced for {\em equilibrated\/}
matter, also to hold in the highly non-equilibrium situation 
present in the experiments, we can state that the higher correlation
radius of the NA49 data might indicate a higher temperature
of the reaction zone rather than its bigger size. 
However, this interpretation depends crucially on the 
momentum range where the pions are measured.
The uncertainties in this statement indicate
the importance of calculations for mesonic spectral functions 
in hot nuclear matter. 

The effect of narrowing the correlation function 
might be even more pronounced due to the short time-scales
of relativistic heavy-ion collisions, which may suppress the spectral
width ($\simeq$ collision rate) of pions even more. Indications for
such a behavior are obtained in simulation codes, where at high enough
collision energies one may reproduce experimental data with
{\em free\/} particle collision rates. Thus, instead of speaking about
``temperature'', which always is connected with the notion of a partial
equilibrium, we may also reformulate our conclusion even more
sharply: Measured correlation functions become narrower due
to non-equilibrium effects in statistically radiating sources.
A bigger correlation radius therefore is an indication, that
a system of colliding heavy-ion collisions is farther from
local equilibrium in the collision zone. 

Finally we point out that in common calculations 
of the correlation function one has to introduce ad-hoc
random phases between several classical sources and then obtains
only the local equilibrium correlator 
$c^{\mbox{\small loc}}_{ll^\prime}(\vec{p},\vec{q})$. Instead, we
relied on a proper field theoretical treatment which incorporates
{\em non-equilibrium effects} correctly up to first order in
gradients of the temperature. We found, that the non-equilibrium character
of the system must be taken serious when calculating the correlation 
function -- which answers the secondary question
we asked in the introduction.

\subsection*{\normalsize\bf Coda}
A large part of the work of E.Wigner was dedicated to the study
of symmetries in physical systems. The fact, that particles
in a thermal state must have a nontrivial spectral function
rather than a sharp ''mass-shell constraint'', is connected
to the breaking of such a symmetry: A thermal state has a
preferred rest frame and therefore violates the Lorentz invariance
of the usual field theoretical ground state \cite{BS75}. 

%
%
\end{document}